# Giant enhancement of higher-order harmonics of an optical-tweezer phonon laser

Guangzong Xiao[1,3,4,*], Tengfang Kuang[1,3,4], Yutong He[1,3], Xinlin Chen[1,3], Wei Xiong[1,3], Xiang Han[1,3], Zhongqi Tan[1,3], Hui Luo[1,3*], Hui Jing[2*]

[1]College of Advanced Interdisciplinary Studies, NUDT, Changsha Hunan, 410073, China

[2]Department of Physics and Synergetic Innovation Center for Quantum Effects and Applications, Hunan Normal University, Changsha Hunan, 410081, China

[3]Nanhu Laser Laboratory, NUDT, Changsha Hunan, 410073, China

[4]These authors contributed equally to this work.

*Corresponding authors: xiaoguangzong@nudt.edu.cn; luohui.luo@163.com; jinghui@hunnu.edu.cn;

**Abstract**

Phonon lasers, as mechanical analogues of optical lasers, are unique tools for not only fundamental studies of phononics but also diverse applications such as acoustic imaging and force sensing. Very recently, by levitating a micro-size sphere in an optical tweezer, higher-order mechanical harmonics were observed in the phonon-lasing regime, as the first step towards nonlinear levitated optomechanics [Nat. Phys. 19, 414 (2023)]. However, both the lasing strengths and the quality factors of the observed harmonics are typically very low, thus severely hindering their applications. Here we show that, by applying a simple but powerful electronic control to such a levitated micro-sphere, three orders of magnitude enhancement are achievable in the brightness of the phonon lasers, including both the fundamental mode and all its higher-order harmonics. Also, giant improvements of their linewidth and frequency stability are realized in such an electro-optomechanical system, together with further improved higher-order phonon coherence. These results, as a significant step forward for enhancing and controlling micro-object phonon lasers, can be readily used for a wide range of applications involving nonlinear phonon lasers, such as acoustic frequency comb, ultra-sound sensing, atmospherical monitoring, and even bio-medical diagnosis of levitated micro-size objects.

# 1. Introduction

The coherent generation and control of phonons are central to a variety of aspects of quantum metrology and quantum information science, including the development of quantum transducers[1–7], solid-state quantum computers[8–10], and metrological techniques for detecting gravitational wave or dark matter[11,12]. Recent developments include the demonstration of acoustic beam splitters[13], nonreciprocal sound devices[14, 15], topological mechanical lattices[16], and vibrational wave mixers[17], to name only a few. Phonon lasers, characterized by a wavelength shorter than optical lasers of the same frequency[18], are expected to play a major role in driving phonon devices and improving force sensors[19], and thus great efforts have been devoted to realizing them with ions[20–23], micro-resonators[24–29], membrane[30,31], semi-conductor lattice[32], and photonic crystal[33]. Recently, a nanosphere phonon laser operating in a levitated optomechanics (LOM) configuration was demonstrated[34], opening the door to exploring phononics with optical tweezers. Then by using a levitated micro-size object, we demonstrated nonlinear multi-colour phonon lasers in an active LOM system, which features simultaneous emerging of weak higher-order mechanical harmonics, in addition to the fundamental-mode phonon laser[35]. Such nonlinear phonon lasers are potentially important for both fundamental studies of nonlinear phononics and diverse applications such as multi-mode phonon sensors or acoustic frequency combs[36].

However, due to strong dissipative couplings of the large-size objects with the light, nonlinear signals of higher-order harmonics in such levitated phonon-laser devices are typically very weak, thus severely hindering their applications in practice. We note that several methods have been presented previously to improve the quality of phonon lasers, such as feedback control[34], optical polarization control[32], and Floquet engineering[37]. Nevertheless, it has remained a highly nontrivial challenge to greatly improve the qualities of both the fundamental-mode phonon laser and all its higher-order harmonics for our levitated micro-sphere system till now. Here, for the first time, we show that by applying a simple but powerful way of electronic injection into our active LOM system, both the fundamental-mode phonon laser and all its higher-order harmonics can be well locked, leading to giant enhancement of their qualities, including brightness, linewidths, frequency stabilities, and higher-order coherence. Our work provides a significant step forward for enhancing and controlling micro-object phonon lasers, which can be highly desirable for a number of applications, especially for ultrasensitive metrology.

We note that injection locking, as a flexible tool to tune all kinds of oscillations[38–43], has previously been used to achieve hertz-linewidth optical lasers[44–46], low-noise soliton microcomb[47–49], and stable mechanical devices[50–53]. In particular, for a trapped-ion phonon laser, electronic injection enabled 2 orders enhancements of both lasing brightness and linewidth narrowing, reaching to a quality factor $Q_m \sim 10^4$ and thus an ultrahigh force sensitivity[54]. Nevertheless, as far as we know, there is no report about simultaneous locking of both fundamental-mode phonon laser and its all higher-order harmonics, no report about giant improvements of nonlinear mechanical harmonics in the phonon-lasing regime, and no report at all about clear evidences of the positive role of locking on enhancing higher-order coherence of phonon lasers.

The new features of our work can be summarized as: (i) Compared to our previous experiment[35], the brightness of the fundamental-mode phonon laser is enhanced by 3 orders of magnitude, with also 5 orders linewidth narrowing, reaching a quality factor $Q_m \sim 6.6 \times 10^6$, i.e., 2 orders higher than those achieved previously with a cold ion[54] or a nanoparticle[55]. (ii) Compared to Ref. 35, the frequency stability of the phonon laser is enhanced for 5 orders of magnitude, leading to a longer trapping lifetime of the micro-object, i.e., from 1.3 minutes to over 1.2 hours. (iii) Giant enhancement can also be observed for all the spontaneously emerging mechanical harmonics. We confirm that this giant enhancement is not merely due to the electronic locking, or merely due to the optical gain in our LOM device, but the constructive interplay of these two key elements. This is actually another new feature of our present work. (iv) For the first time, we also identify the positive role of locking in enhancing the higher-order coherence of phonon lasers, as measured by higher-order phonon correlations. (v) Mode splitting is observed for our phonon lasers when injecting a strong electronic signal. With these new features and advantages, we believe our active LOM system can serve as powerful new tools to explore levitated nonlinear optomechanics with different types of large-mass objects, for diverse phonon-involving applications.

## 2. Results

The experimental system is illustrated in Fig. 1a. It consists of a dual-beam optical tweezer, a $SiO_2$ microsphere with intrinsic frequency $\Omega_0 = 10.1$ kHz and an active cavity driven by a CW laser of power $P$[35]. This active LOM system is electronically driven by an electrode along the $x$-axis below the trapped microsphere and coupled to a signal generator.

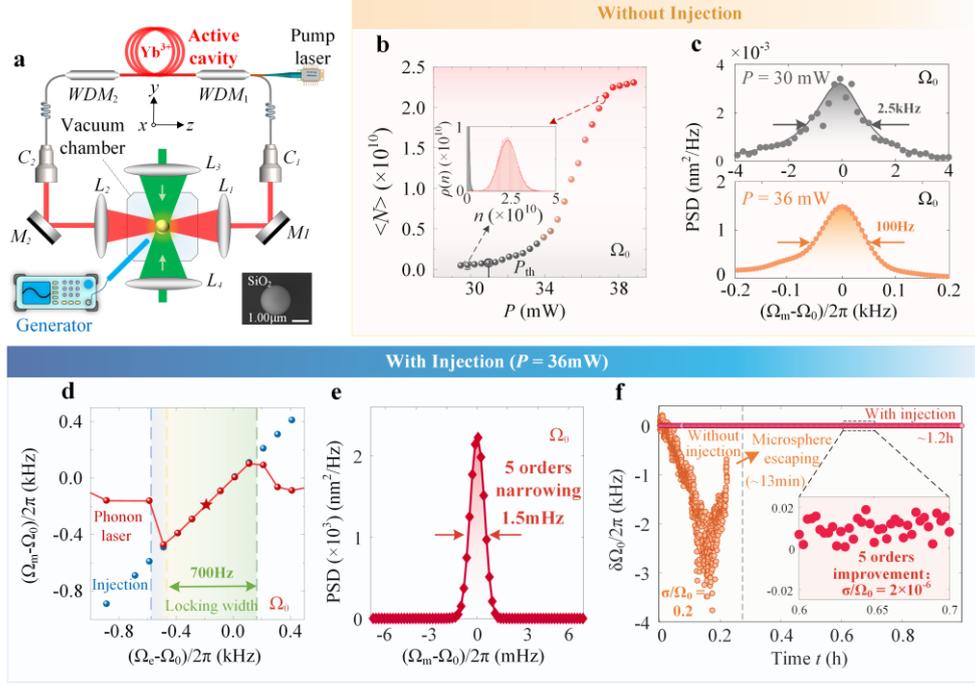

**Fig. 1 Electronically-driven active LOM system with a micro-size sphere. a.** Schematic diagram of the experimental setup, including an active cavity (red), a dual-beam optical tweezer (green), and an electrode (blue). $L_i$, Lenses; $C_i$, collimators; $WDM_i$, wavelength division multiplexers; $M_i$, reflectors. **b.** Phonon population $\langle N \rangle$ of the fundamental-mode phonon laser of frequency $\Omega_0$ as a function of the laser pump power $P$. Inset: Phonon probability distributions below (grey) and above (coloured) the power threshold $P_{th}$. **c.** Power spectral density (PSD) of the fundamental phonon laser mode below (upper curve) or above (lower curve) the threshold. **d.** Measured phonon frequency as a function of the electronic frequency $\Omega_e$, plotted over the locking width. **e.** PSD of the phonon laser when $\Omega_e$ falls within the locking width (marked in **d**), showing a linewidth narrowing of 5 orders in comparison to the case without any injection, as shown in **c**. **f.** Comparison of mechanical frequency stabilities for the unlocked case (orange) and locked case (red). The inset shows a detailed view of the corresponding curve. $\sigma/\Omega_0$: Relative standard deviation, where $\sigma$ is the standard deviation of the mechanical frequency.

A tunable alternating current (AC) field of frequency $\Omega_e$ and amplitude $U_e$, is applied to drive the microsphere. The distance d between the tip of the electrode and the trap centre is $\sim$ 3 mm, and the natural charge $q_e$ of the microsphere was measured as $2.35 \times 10^4$ e, estimated by $q_e = \langle F_{el} \rangle / E_e$, with[56]

$$F_{el} = \sqrt{S_x^{el}(\Omega) / \frac{2\tau \sin c^2[2(\Omega - \Omega_e)\tau]}{M^2[(\Omega^2 - \Omega_0^2)^2 + \Gamma_x^2 \Omega^2]}} \quad (1)$$

where $F_{el}$ is the electric force, $E_e = U_e/d$ the electric field intensity, $M$ the mass of the microsphere, $\Gamma_x$ the mechanical damping rate, $\tau$ the measurement time, and $S_x^{el}(\Omega)$ the power spectral density (PSD) value at the frequency $\Omega_e$.

We first consider the case without any electronic control, as reported in Ref. 35. As shown in Fig. 1b, by increasing the pump power $P$ of the active cavity, the microsphere can be driven through a threshold from the thermal to the coherent phonon lasing regime, as also confirmed by phonon correlation function measurements discussed later in this paper. The inset of Fig. 1b shows the clear change of the phonon number distribution $p(n)$ from Boltzmann to the Gaussian. As also observed in the tweezer phonon laser of Ref. 34, we find that the variance of the distribution is clearly smaller than that of a thermal state with the same mean phonon number, a situation referred to as subthermal number squeezing, with a higher degree of squeezing leading to Poissonian statistics[34,35]. The threshold of the fundamental phonon laser at $\Omega_0$ is clearly apparent in the mean phonon population $<N> = M\Omega_0 x^2/\hbar$, with $x$ its center-of-mass displacement. Also, we observe that, below the threshold, the phonon mode linewidth is 2.5 kHz (for $P$ = 30 mW), and it is reduced to about 100 Hz for $P$ = 36 mW above the threshold (see Fig 1c).

With these results at hand, we now turn on the electronic control and scan $\Omega_e$ for a fixed $U_e$ = 0.9 V. The electronic force on the sphere is ~ 4.47×10$^{-12}$ N, with a displacement amplitude of ~52 nm. The resulting shift $\Omega_m$ - $\Omega_0$, plotted in Fig. 1d, clearly shows the locking of the phonon-laser frequency to $\Omega_e$. The locking width of this experiment is 700 Hz, which can be well described by the theoretical mode[54]

$$\omega_m = \frac{q_e E_e}{2M\Omega_0 x_0},\qquad(2)$$

where $x_0$ is the amplitude of the microsphere. When compared with the unlocked system of Fig. 1c[35], we observe a 3 order of magnitude increase of the PSD and a 5 orders linewidth narrowing for the phonon laser, see Fig. 1e, with the vibrating amplitude of the microsphere increased from ~ 127 nm to ~ 146 nm. Correspondingly, the mechanical quality factor reaches 6.6×10$^6$. Higher mechanical quality factor was indeed achievable for a levitated nanosphere, using sophisticated feedback controls based on electronic loops[57,58]. However, such a technique is difficult to apply for a micro-size sphere, due to much stronger scattering losses[59]. In our previous work, we introduced an optical gain to overcome this obstacle and achieve a microsphere phonon laser with a quality factor of ~ 400[35]. In comparison to Ref. 35, the mechanical quality factor here is improved for 4 orders, which is actually the highest record as far as we know, for a micro-sphere phonon laser.

Also, we find that the phonon-laser's frequency stability can be significantly improved, as shown in Fig. 1f. In the absence of any electronic control, the mechanical frequency drifts due to thermal noises and gain-induced heating[60], with a standard deviation of $\sigma \approx$ 1886.7 Hz. This eventually leads to the escape of the sphere from the trap after ~ 13 min (see the orange dots). In contrast, for the present electronically-controlled system, the trapping lifetime of the microsphere can be increased to more than 1.2 hours, with the standard deviation of the phonon frequency sharply falling down to $\sigma \approx 0.022$ Hz. That is, the relative standard deviation $\sigma/\Omega_0$ is improved for 5 orders of magnitude, from 0.2 to $2\times 10^{-6}$, a significant improvement compared to that achieved in Ref. 35.

We emphasize that in contrast with previous works on injection locking of levitated oscillators[54,55], in our present system, both the electronic injection and the optical gain are important. The giant enhancement of phonon lasers, as reported here, is NOT merely due to the electronic locking. In fact, as demonstrated in Ref. 35, no significant multimode phonon lasing can be observed for a micro-size object in a passive LOM system. In contrast, clear evidence of multimode lasing is observed with an active cavity, a consequence of the gain-enhanced dissipative optomechanical coupling[35]. This fact is here confirmed by a series of measurements that apply only the injection locking to the passive LOM system. This resulted in a weak peak at the locking frequency $\Omega_e$, but none at the mechanical harmonics $n\Omega_e$, $n \geq 2$, see Fig. 2b, demonstrating that the giant enhancement of phonon laser is NOT merely due to the electronic locking. On the other hand, of course, the giant enhancement of phonon laser is also important to be observed with merely the optical gain. In fact, the key to significantly enhancing the phonon lasers is the constructive interplay of these two important factors, which is not straightforward to see at all.

Indeed, this constructive interplay can lead to not only the giant enhancement of the fundamental-mode phonon laser, but also the giant enhancement of all its higher-order harmonics. As shown in the grey curves of Fig. 2a, higher-order harmonics emerge spontaneously due to gain-enhanced LOM nonlinearity[35]. By applying the electric control, these harmonics can also be locked, resulting in 3 to 4 orders enhancement in the PSD. This is clearly different from a previous experiment in which a signal close to mechanical harmonic or subharmonic frequencies was injected into a microtoroidal optomechanical oscillator[50] and then resulted in only the locking of the fundamental mode. As such, our results represent a significant advance from that work since here in the nonlinear regime, both the fundamental phonon-laser mode and all its higher-order harmonics can be simultaneously locked, resulting in giant enhancement of otherwise very weak nonlinear harmonics, by applying only a single-color signal. Also, we remark that our work is clearly different from the recent one on microcavity phonoritons[61], which applied an electronic signal to an acoustic resonator to generate driving phonons, with strict frequency-matching conditions in their dispersive system.

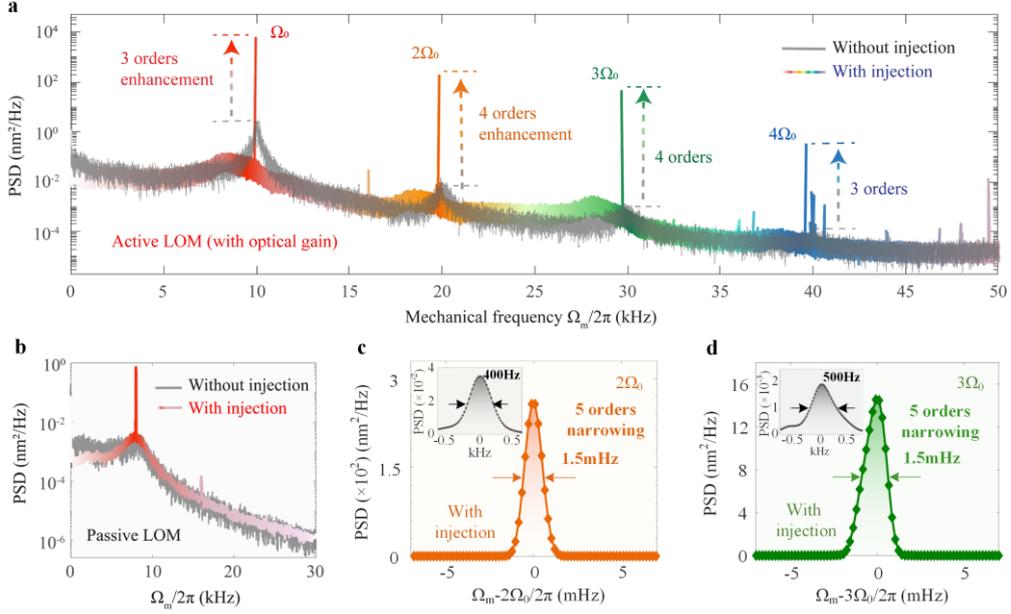

**Fig. 2 Giant enhancement of nonlinear phonon lasers, especially the spontaneously emerging higher-order harmonics. a.** The measured PSD of phonons in the active LOM system for the case without (grey) or with (coloured) the injection. **b.** The PSD for a passive LOM system. **c, d.** The detailed PSDs of the double-frequency mode $2\Omega_0$ and the triple-frequency mode $3\Omega_0$. Insets: The PSDs for the case without any electronic control.

Importantly, as shown in Figs. 2c and 2d the linewidths of all mechanical harmonics of the phonon laser are also very significantly improved by the injection locking: In the absence of any injection, the linewidths of the second and third order harmonics at $2\Omega_0$ or $3\Omega_0$ are about 400 Hz and 500 Hz, respectively, and the electronic injection, reduces them by about 5 orders of magnitude, corresponding to mechanical quality factors of $1.32 \times 10^7$ and $1.98 \times 10^7$, respectively.

In order to further confirm the coherent nature of phonon modes in our present system, we carried out to reveal the locking impact on the higher-order coherence of the phonon lasers by measuring the second to fourth-order phonon field autocorrelations functions at zero-time delay[62], $g^{(k)}(0)=\langle\hat{b}^{\dagger k}\hat{b}^{k}\rangle/\langle\hat{b}^{\dagger}\hat{b}\rangle^{k}$, $k = 2, 3, 4$, with $\hat{b}$ and $\hat{b}^{\dagger}$ the annihilation and creation operators of the mechanical mode, with

$$g^{(2)}(0) = \frac{\langle\hat{N}^2\rangle - \langle\hat{N}\rangle}{\langle\hat{N}\rangle^2}, \qquad (3)$$

$$g^{(3)}(0) = \frac{1}{\langle\hat{N}\rangle^3}(\langle\hat{N}^3\rangle - 3\langle\hat{N}^2\rangle + 2\langle\hat{N}\rangle), \qquad (4)$$

$$g^{(4)}(0) = \frac{1}{\langle\hat{N}\rangle^4}(\langle\hat{N}^4\rangle - 6\langle\hat{N}^3\rangle + 11\langle\hat{N}^2\rangle - 5\langle\hat{N}\rangle), \qquad (5)$$

and $\hat{N}$ is the mean phonon number in the mode under consideration.

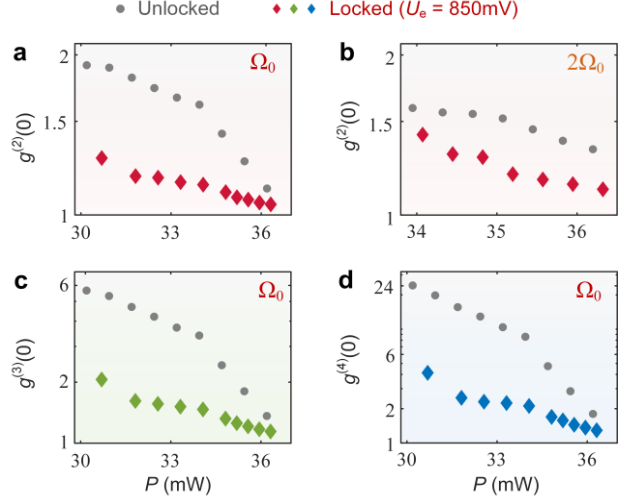

**Fig. 3 Phonon correlations for unlocked (grey) and locked (coloured) LOM systems. a.** Second-order autocorrelations at zero-time delay, $g^{(2)}(0)$, versus pump power $P$ for $\Omega_0$. **b.** $g^{(2)}(0)$ for $2\Omega_0$. **c, d.** Third or fourth order autocorrelations, $g^{(3)}(0)$ and $g^{(4)}(0)$, for $\Omega_0$.

In the absence of electronic locking, as the pump power $P$ is increased past the laser threshold, the second-order correlation functions $g^{(2)}(0)$ of both the fundamental and first harmonic modes decrease to approach 1, see the grey dots in Figs. 3a and 3b. The initial value $g^{(2)}(0)$ of $2\Omega_0$ is lower than 2 since the lasing threshold of $2\Omega_0$ is lower than $\Omega_0$ (see also Ref. [35]). However, $g^{(2)}(0)$ fails to reach 1 due to thermal noise[35] In contrast, when applying the injection signal (red dots on the figures) $g^{(2)}(0)$ approaches $g^{(2)}(0) = 1$ more rapidly, with a lower value than without locking for a given optical driving strength. Similar results are also observed for higher order phonon autocorrelations $g^{(3)}(0)$ and $g^{(4)}(0)$ of the fundamental mode $\Omega_0$, see Figs. 3c and 3d. These results, to the best of our knowledge, have not been reported so far for phonon lasers, and clearly evidence the positive role of injection locking in improving both their second-order and higher-order coherence. We expect that our results can stimulate more works on electronic control of quantum effects of various phonon lasers in different systems ranging from cold ions to membranes and semiconductor lattices[20–33].

As a final series of measurements, we explore the influence of electronic amplitudes on phonon lasers. As shown in Fig. 4a, when tuning the injection from 0 to 900 mV, the fundamental-mode phonon population gradually increases, and similar features are also confirmed for the higher-order harmonics (not shown here). More interestingly, mode splitting emerges for a stronger pump, see Fig. 4b-c. The underlying mechanism and its relation with feedback-induced singularities as revealed in recent works[63,64], a very interesting topic but well beyond the scope of the present work, will be studied in detail in our future works.

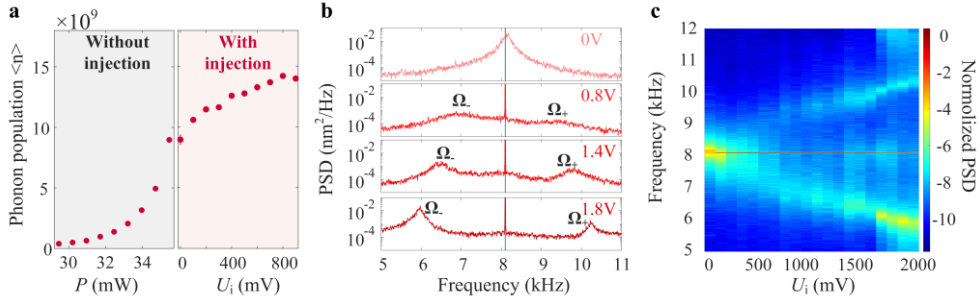

**Fig. 4 Characteristic of the phonon laser for an enhanced injection voltage. a.** Phonon population <n> of the fundamental modes $\Omega_0$ versus the pumping power $P$ and the injection voltage $U_i$. **b.** Mode splitting of the fundamental modes $\Omega_0$, shown with four typical injection voltage $U_i$. **c.** The measured power spectral density as a function of injection voltage $U_i$.

## 3 Discussion

In summary, by the direct application of electronic injection locking into an active LOM system, we have demonstrated a considerable improvement of several key features of multi-mode phonon lasers with micro-size spheres. We found that an effective locking of both the fundamental mode and its higher-order mechanical harmonics is achieved, with three orders of magnitude brighter and five orders of magnitude narrower linewidth, corresponding to mechanical quality factors of more than $Q_m = 10^6$. We also demonstrated that, for the same optical driving strength, the electronic injection locking results in the second and higher correlation functions of the phonon field approaching unity past the laser threshold more rapidly than the case without locking. Mode splitting is also observed by further increasing the injection amplitude.

As the final remark, we know that in previous experiments[51,55], the locking signals were generated optically and transformed into electronic signals through an electro-to-optical modulator approach. In contrast, in our approach, the electronic injection signal is applied directly from an electrode, without any need of additional electro-to-optical conversion devices, considerably reducing the rigorous requirements on injection sources and noise from diverse devices. Our simplified approach results in the highest quality factor of the fundamental mode phonon laser reported so far, and also enables a giant enhancement in the intensity of the higher-order harmonic phonon modes. This hybrid electro-LOM system offers considerable promise as a platform to explore novel applications of coherent electro-to-acoustic conversions[10,65], nonlinear phononics, and the development of precise acoustic sensors. For example, inspired by the very recent experiment on phononic frequency comb[36], we expect that our system can be used to achieve such a frequency comb by levitating highly nonlinear objects or adding tunable nonlinear materials into the cavity. Future works will also further investigate the underlying mechanisms of multimode phonon lasers, explore other possibilities such as achieving squeezed or correlated phonon lasers[66], and consider their applications in quantum metrology, in particular in the development of high-precision force sensors[66].

## 4 Materials and methods

**Experimental details**

The experimental setup consisted of four parts, including an active cavity, a dual-beam optical tweezer, electronic injection devices and position detection system. As shown in Fig. 1a, the active cavity was vertical to the dual-beam optical tweezer around the trapping region. The active cavity was composed of a fiber laser path (with optical gain) and a free-space laser path. We installed the free-space laser path of the active cavity to a 3D translation stage. Then, the relative position of the active cavity to the trapped microsphere was tuned with resolution of 0.1 μm. Detailed parameter was also found in Ref. 35. We applied electronic injection to the trapped microsphere through an electrode, which was installed directly below the microsphere in the $x$ direction. It was made of polished stainless steel with a tip radius of 200μm. The distance $d_0$ between the trap centre and the electrode is measured to be about 3 mm. A function signal generator (DG1402-8750) is installed to generate the electronic field, then a tunable alternating current (AC) field could be applied to the microsphere.

**Experimental procedure**

The experiment was mainly operated with the following three steps. (i) Microsphere trapping. Microspheres were loaded into the trapping region with a nebulizer. In most cases, a microsphere was trapped within 30 s. Once a microsphere was trapped, we switched on the vacuum pump system and reduced the pressure to the desired level. (ii) Phonon lasing. The free-space laser path of the active cavity is installed on a 3D translation stage. We adjusted its relative position to the microsphere in order to optimize the dissipative coupling strength. The phonon lasing behavior was observed, according to the real time PSD of the microsphere's displacement. (iii) Electronic injection and data acquisition. We adjusted the amplitude and frequency of the signal generator according to the phonon lasing dynamic. Then, we investigated the influence of the electronic injection on the characteristic of the fundamental mode phonon laser and it high-order harmonics, as shown in the main text.

**Charge estimation**

The natural charge of $q_e$ of the microsphere was estimated by $q_e = \langle F_{el} \rangle / E_e$, with the relationship $F_{el} = \sqrt{S_x^{el}(\Omega) / \frac{2\tau \sin c^2[2(\Omega - \Omega_e)\tau]}{M^2[(\Omega^2 - \Omega_0^2)^2 + \Gamma_x^2 \Omega^2]}}$. The equation came from the Langevin equation of the microsphere, which could be expressed as $m\ddot{x} + m\Gamma_x \dot{x} + kx = F_{th}(x) + F_{el}(t)$, where $F_{el}(t) = F_{el-x} \cos(\omega_{dr} t)$. Then, the power density of the microsphere could be calculated as $S_x^{el}(\omega) = \frac{S_{vx}^{el}(\omega)}{\beta^2} = \frac{2 F_{el-x}^2 \tau \sin c^2[2(\omega - \omega_{dr})\tau]}{m^2[(\omega^2 - \omega_x^2)^2 + \Gamma_x^2 \omega^2]}$. In the experiment, we got the PSD from the displacement of the microsphere. As the sphere altered the spatial distribution of the scattering light, the differential mode signal from the BPD revealed the sphere's position. With the electric force in hand, we can estimate the natural charge $q_e$. Here, we approximated the electric field strength to the strength of the input electrical signal, and then for a single suspended microsphere the charge can be calculated.